\documentclass[sigconf]{acmart}

\AtBeginDocument{%
  \providecommand\BibTeX{{%
    \normalfont B\kern-0.5em{\scshape i\kern-0.25em b}\kern-0.8em\TeX}}}

\usepackage{amsmath,amsfonts}
\usepackage{algorithm}
\usepackage{algorithmic}
\usepackage{graphicx}
\usepackage{textcomp}
\usepackage{xcolor}
\usepackage{booktabs}
\usepackage{multirow}
\usepackage[super]{nth}
\usepackage{tikz}
\newcommand*\circled[1]{\tikz[baseline=(char.base)]{
		\node[shape=circle,draw,inner sep=0.2pt] (char) {#1};}}
\newcommand*\circledB[1]{\tikz[baseline=(char.base)]{
            \node[shape=circle,fill,inner sep=0.2pt] (char) {\textcolor{white}{#1}};}}
\usepackage{soul}
\usepackage{enumitem}
\usepackage{setspace}

\usepackage{hyphenat}
\usepackage{fancyhdr}
\fancypagestyle{firstpage}{
  \fancyhf{}
  \chead{\small To appear at the 59th IEEE/ACM Design Automation Conference (DAC), July 2022, San Francisco, CA, USA.}
  \cfoot{\thepage}
}

\begin{document}

\title{SoftSNN: Low-Cost Fault Tolerance for Spiking Neural Network Accelerators under Soft Errors}

% \affiliation{%

\author{Rachmad Vidya Wicaksana Putra$^*$, Muhammad Abdullah Hanif$^{\dagger}$, Muhammad Shafique$^\dagger$}
\affiliation{%
  \institution{$^*$Technische Universit\"at Wien (TU Wien) \city{Vienna} \country{Austria}}
  \institution{$^\dagger$New York University Abu Dhabi (NYUAD) \city{Abu Dhabi} \country{United Arab Emirates}}}
\email{rachmad.putra@tuwien.ac.at, {mh6117, muhammad.shafique}@nyu.edu}


%% The abstract is a short summary of the work to be presented in the article.
\begin{abstract}
\begin{spacing}{0.91}
Specialized hardware accelerators have been designed and employed to maximize the performance efficiency of Spiking Neural Networks (SNNs). 
However, such accelerators are vulnerable to transient faults (i.e., soft errors), which occur due to high-energy particle strikes, and manifest as bit flips at the hardware layer. 
These errors can change the weight values and neuron operations in the compute engine of SNN accelerators, thereby leading to incorrect outputs and accuracy degradation.
However, the impact of soft errors in the compute engine and the respective mitigation techniques have not been thoroughly studied yet for SNNs. 
A potential solution is employing redundant executions (re-execution) for ensuring correct outputs, but it leads to huge latency and energy overheads.
Toward this, we propose \textit{SoftSNN}, a novel methodology to mitigate soft errors in the weight registers (synapses) and neurons of SNN accelerators \textit{without re-execution}, thereby maintaining the accuracy with low latency and energy overheads. 
Our SoftSNN methodology employs the following key steps: (1) analyzing the SNN characteristics under soft errors to identify faulty weights and neuron operations, which are required for recognizing faulty SNN behavior; (2) a Bound-and-Protect technique that leverages this analysis to improve the SNN fault tolerance by bounding the weight values and protecting the neurons from faulty operations; and (3) devising lightweight hardware enhancements for the neural hardware accelerator to efficiently support the proposed technique. 
The experimental results show that, for a 900-neuron network with even a high fault rate, our SoftSNN maintains the accuracy degradation below 3\%, while reducing latency and energy by up to 3x and 2.3x respectively, as compared to the re-execution technique.
\end{spacing}
\end{abstract}



% To remove the ACM Reference Format Example 
\settopmatter{printacmref=false} % Removes citation information below the abstract
\renewcommand \footnotetextcopyrightpermission[1]{} % Removes footnote with conference information in first column

% \keywords{Spiking neural networks, SNNs, accelerators, compute engine, transient faults, soft errors, reliability, fault tolerance, low overheads.}

\maketitle
\pagestyle{plain}
\thispagestyle{firstpage}

\begin{spacing}{0.91}
%%%%%%%%%%%%%%%%%%%%%%%%%%%%%%%%%%%%%%%%%%%%%%%%%%
%%%%%%%%%%%%%%%%%%%%%%%%%%%%%%%%%%%%%%%%%%%%%%%%%%

\vspace{-0.2cm}
\section{Introduction}
\label{Sec_Intro}

\begin{figure}[t]
\centering
\includegraphics[width=0.92\linewidth]{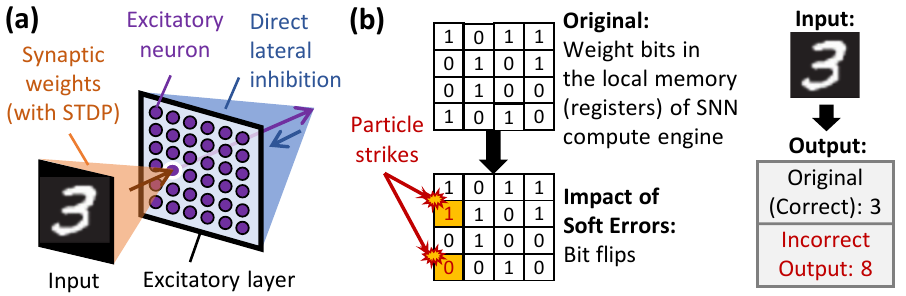}
\vspace{-0.4cm}
\caption{(a) The SNN architecture considered in this work (i.e., a fully-connected network). (b) High-energy particle strikes trigger soft errors as bit flips in the hardware layer (e.g., the weight registers of compute engine), and result in incorrect output (e.g., misclassification) in application layer.}
\label{Fig_SNN_SoftErrors}
\vspace{-0.6cm}
\end{figure}

\begin{figure}[t]
\centering
\includegraphics[width=\linewidth]{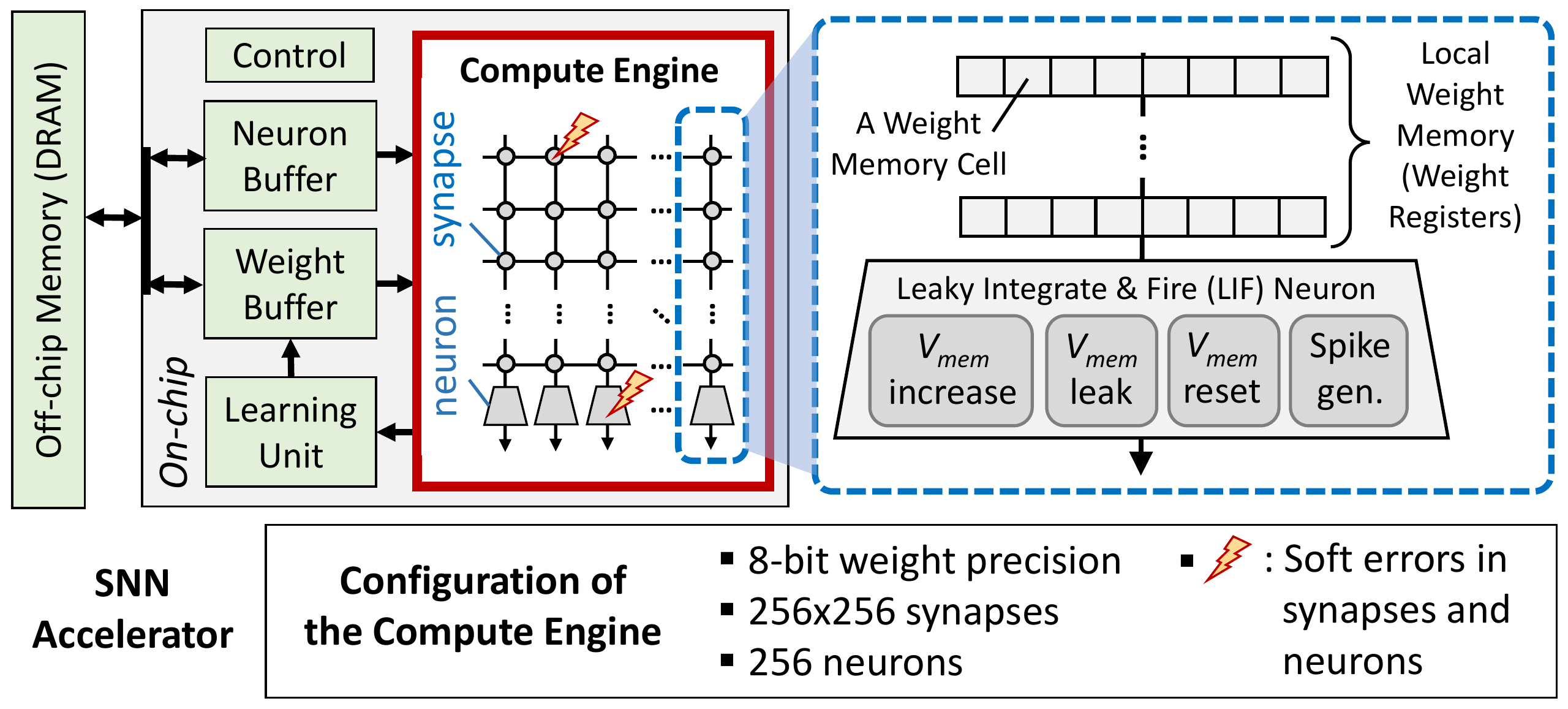}
\vspace{-0.7cm}
\caption{The typical architecture of SNN accelerators. 
Detailed discussion on this architecture is provided in Section~\ref{Sec_Prelim_SNNs}, and the transient fault modeling is in Section~\ref{Sec_Prelim_FaultModel}.}
\label{Fig_SNNacc_Overview}
\vspace{-0.2cm}
\end{figure}

SNNs have shown a great potential for obtaining high accuracy in classification tasks (e.g., digit classification, object recognition, etc.) with very low processing power/energy due to their sparse and event-based operations of spiking activity~\cite{Ref_Putra_FSpiNN_TCAD20}. 
Moreover, SNNs can perform unsupervised learning using unlabeled data due to their bio-plausible learning mechanism like the spike-timing-dependent plasticity (STDP); see an SNN architecture that supports unsupervised learning in Fig.~\ref{Fig_SNN_SoftErrors}(a).
This characteristic is needed in several real-world autonomous systems (e.g., robotics and UAVs) where collecting unlabeled data is easier and cheaper than labeled data~\cite{Ref_Putra_FSpiNN_TCAD20}.
To achieve these benefits while maximizing the performance efficiency of SNN processing, specialized hardware accelerators have been proposed in the literature~\cite{Ref_Painkras_SpiNNaker_JSCC13, Ref_Akopyan_TrueNorth_TCAD15, Ref_Davies_Loihi_MM18, Ref_Frenkel_ODIN_TBCAS19, Ref_Chen_SNNchip_TCSII21}.
However, such hardware accelerators are vulnerable to transient faults (i.e., soft errors)~\cite{Ref_Baumann_SoftErrors_DnT05} due to high-energy particle strikes, which may come from cosmic rays or packaging materials. 
These errors manifest as bit flips at the hardware layer, and can propagate to the application layer, resulting in incorrect outputs~\cite{Ref_Baumann_SoftErrors_DnT05}, as shown in Fig.~\ref{Fig_SNN_SoftErrors}(b). 
Soft errors in the local weight memory/registers and the neurons can affect the functionality of the SNN compute engine, e.g., by corrupting the weight values and the behavior of neuron operations including operations for membrane potential dynamics and spike generation (see Fig.~\ref{Fig_SNNacc_Overview}). 
These conditions may lead to significant accuracy degradation in the inference phase (which will be discussed further in Section \ref{Sec_SoftSNN_FTAnalysis}.), thereby posing reliability threats for SNN-based systems. 
It is especially important when such systems are used for safety-critical applications (e.g., medical data analysis).

\textit{\textbf{Targeted Research Problem:} If and how can we efficiently mitigate the negative impact of soft errors in the SNN compute engine\footnote{For conciseness, we use ``SNN compute engine" or ``compute engine" interchangeably to denote the compute engine of an SNN accelerator.} (including both the local weight registers and the neurons) on accuracy at run time, while avoiding the excessive overhead of redundant executions.}
An efficient solution to this problem will enable reliable SNN executions under soft errors with low latency and energy overheads, thereby making it suitable for latency- and energy-constrained applications (like IoT-Edge).

%%%%%%%%%%%%%%%%%%%%%%
\vspace{-0.2cm}
\subsection{State-of-the-Art and Their Limitations}
\label{Sec_Intro_SOA}

To mitigate soft errors, the conventional fault tolerance techniques for VLSI circuits, like Error Correction Code (ECC)~\cite{Ref_Sze_ECCs_USPatent00}, Dual Modular Redundancy (DMR)~\cite{Ref_Vadlamani_DMR_DATE10}, and Triple Modular Redundancy (TMR)~\cite{Ref_Lyons_TMR_IBM62}, can be employed.
However, they require extra/redundant executions and/or hardware, which incur huge area and energy overheads for correcting a limited number of faulty bits in memories. 
State-of-the-art works have studied different design aspects for understanding and improving SNN fault tolerance, which can be classified into the following categories. 

\begin{itemize}[leftmargin=*]
    \item \textbf{Fault Modeling:} 
    Possible faults in SNN components (i.e., neurons and synapses) have been identified in~\cite{Ref_Vatajelu_ReliabilitySNN_VTS19}. 
    In the analog domain, fault modeling for analog neuron hardware has been discussed in \cite{Ref_Sayed_NeuronFaultModel_IOLTS20,Ref_Spyrou_NeuronFT_DATE21}. Unlike this, we target SNN implementation in digital domain requiring a different approach. 
    \item \textbf{Fault Tolerance Studies:} Previous works have performed fault injection in form of synaptic failure/removal on the SNN weights without considering the underlying hardware architectures and processing dataflows~\cite{Ref_Schuman_ResilinceSNN_IJCNN20, Ref_Rastogi_AstrocytesSTDP_FNINS21}. 
    Recent works~\cite{Ref_Putra_SparkXD_DAC21,Ref_Putra_ReSpawn_ICCAD21} have proposed techniques for mitigating approximation-induced errors and permanent faults, which reside in the weight memories (i.e., DRAM and weight buffer) of SNN accelerators.
\end{itemize}
The above-discussed state-of-the-art techniques have not studied the SNN fault tolerance considering soft errors in the underlying hardware.
Therefore, \textit{the impact of soft errors in the compute engine (i.e., local weight registers and neurons)\footnote{We focus on the SNN compute engine, as it is responsible for generating spikes which determine the classification activity, thereby having a significant impact on accuracy.} on the accuracy, and the respective lightweight mitigation techniques are still unexplored}. 
To identify and better understand the challenges for mitigating soft errors in SNN accelerators,
we first perform an experimental case study in Section~\ref{Sec_Intro_CaseStudy}.

\begin{figure}[t]
\vspace{-0.2cm}
\centering
\includegraphics[width=\linewidth]{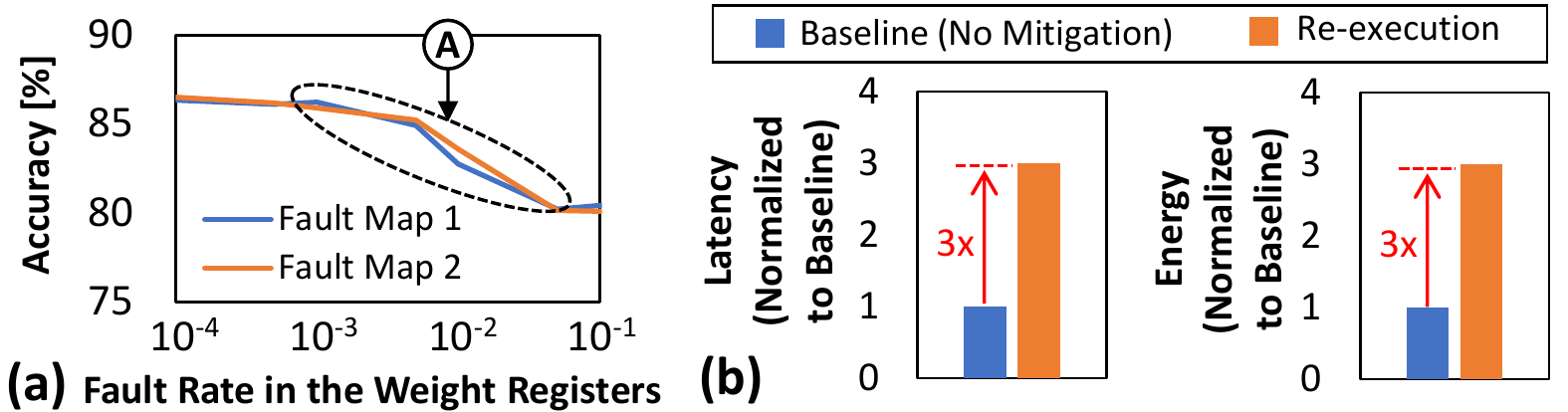}
\vspace{-0.7cm}
\caption{Results of a 400-neuron network on the MNIST for (a) accuracy, considering different fault locations (fault maps) and fault rates in the weight registers of the compute engine, and (b) latency and energy for different designs.}
\label{Fig_Observe_FaultyWreg}
\vspace{-0.5cm}
\end{figure}

%%%%%%%%%%%%%%%%%%%%%%
\vspace{-0.2cm}
\subsection{Case Study and Research Challenges}
\label{Sec_Intro_CaseStudy}

We study the impact of faulty weight registers on accuracy, while considering the SNN architecture with direct lateral inhibition and STDP learning as shown in Fig.~\ref{Fig_SNN_SoftErrors}(a) and the hardware architecture shown in Fig.~\ref{Fig_SNNacc_Overview}.
Due to the criticality of soft errors in the memory components that are built with relatively small-sized transistors, as compared to other parts of the circuit, we first perform fault injection experiments on the weight registers of different neurons with random distribution, while considering different fault maps and different fault rates. Details on the experimental setup are provided in Section~\ref{Sec_Evaluation}.
The experimental results are shown in Fig.~\ref{Fig_Observe_FaultyWreg}, from which we derive the following key observations.

\begin{itemize}[leftmargin=*]
    \item  Different combinations of fault maps and fault rates (which represent different possible soft error patterns in real-world conditions) lead to diverse accuracy profiles, even for the same SNN model and workload, indicating its unpredictable nature at design time; see \circled{A} in Fig.~\ref{Fig_Observe_FaultyWreg}(a).
    \item A potential solution is to employ redundant executions (i.e., re-execution) as it does not require hardware modification, but at the cost of huge latency and energy overheads; see Fig.~\ref{Fig_Observe_FaultyWreg}(b). 
\end{itemize}

Based on these observations, we highlight the following \textit{research challenges} in devising solutions for the targeted problem.
\begin{itemize}[leftmargin=*]
    \item \textit{The mitigation technique should recognize any faulty components (weight registers and neurons operations) at run time}, to cope with unpredictable run-time scenarios of different soft error profiles.
    \item \textit{The mitigation should not employ re-execution}, because it requires huge latency and energy overheads.
    \item \textit{The mitigation should have minimal latency and energy overheads} compared to that of the ``SNN without mitigation'', thereby making it applicable for latency- and energy-constrained applications.
\end{itemize}

%%%%%%%%%%%%%%%%%%%%%%
\renewcommand{\headrulewidth}{0pt}
\vspace{-0.2cm}
\subsection{Our Novel Contributions}
\label{Sec_Intro_NovelContrib}

\begin{figure}[hbtp]
\vspace{-0.3cm}
\centering
\includegraphics[width=\linewidth]{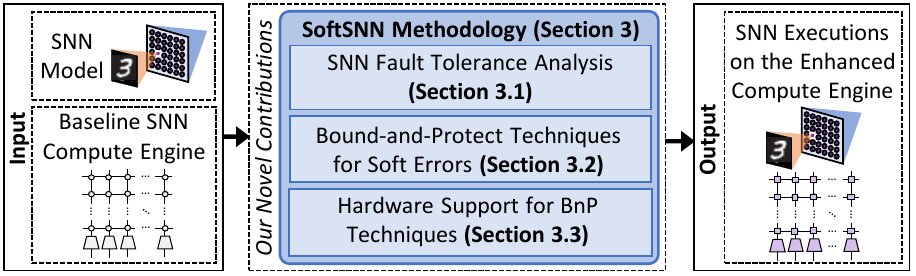}
\vspace{-0.8cm}
\caption{Overview of our novel contributions.}
\label{Fig_SoftSNN_Novelty}
\vspace{-0.4cm}
\end{figure}

To address the above challenges, we propose \textbf{SoftSNN}, \textit{a novel methodology that enables reliable SNN processing on hardware accelerators under soft errors without re-execution}. 
To the best of our knowledge, this work is the first effort that studies the impact of soft errors on SNN execution on the SNN accelerators, and develops a cost-effective mitigation technique.
Our SoftSNN methodology employs the following key steps (see overview in Fig.~\ref{Fig_SoftSNN_Novelty}).
\begin{itemize}[leftmargin=*]
    \item \textbf{Analyzing the SNN fault tolerance under soft errors} to understand the impact of faulty SNN components (i.e., synapses and neurons) on accuracy under different fault rates. 
    \item \textbf{Employing different Bound-and-Protect (BnP) techniques} to bound the weight values within a safe range, and protect the neurons from performing faulty operations, based on the information from SNN fault tolerance analysis.
    \item \textbf{Employing lightweight hardware support for the BnP techniques} to efficiently identify when faulty weight values and neuron operations occur, then perform weight bounding and generate safe neuron behavior that does not cause significant accuracy degradation.  
\end{itemize}

\textbf{Key Results:} 
We evaluate our SoftSNN methodology using a Python-based framework on a multi-GPU machine. 
We perform 3 epochs of unsupervised training (3x60K experiments) for each combination of SNN model and workload; and 10K experiments for each combination of SNN model, workload, and fault rate for inference.
The experimental results show that, compared to the re-execution, the SoftSNN keeps the accuracy degradation below 3\%, while reducing up to 3x and 2.3x of latency and energy respectively, for a 900-neuron SNN with a 0.1 fault rate and the MNIST workload.  
 
%%%%%%%%%%%%%%%%%%%%%%%%%%%%%%%%%%%%%%%%%%%%%%%%%%
%%%%%%%%%%%%%%%%%%%%%%%%%%%%%%%%%%%%%%%%%%%%%%%%%%
\vspace{-0.2cm}
\section{Preliminaries}
\label{Sec_Prelim}

%%%%%%%%%%%%%%%%%%%%%%
\subsection{Spiking Neural Networks (SNNs)}
\label{Sec_Prelim_SNNs}

\textbf{Overview:}
In an SNN model, neurons and synapses are connected in a specific fashion that defines the network architecture.
In this work, we consider a fully connected architecture with direct lateral inhibition and STDP learning, as shown in Fig.~\ref{Fig_SNN_SoftErrors}(a), because it can achieve high accuracy in an unsupervised scenario~\cite{Ref_Putra_FSpiNN_TCAD20}.
For the neuron model, we employ the Leaky Integrate and Fire (LIF), due to its low computational complexity. 

\begin{figure}[hbtp]
\vspace{-0.2cm}
\centering
\includegraphics[width=\linewidth]{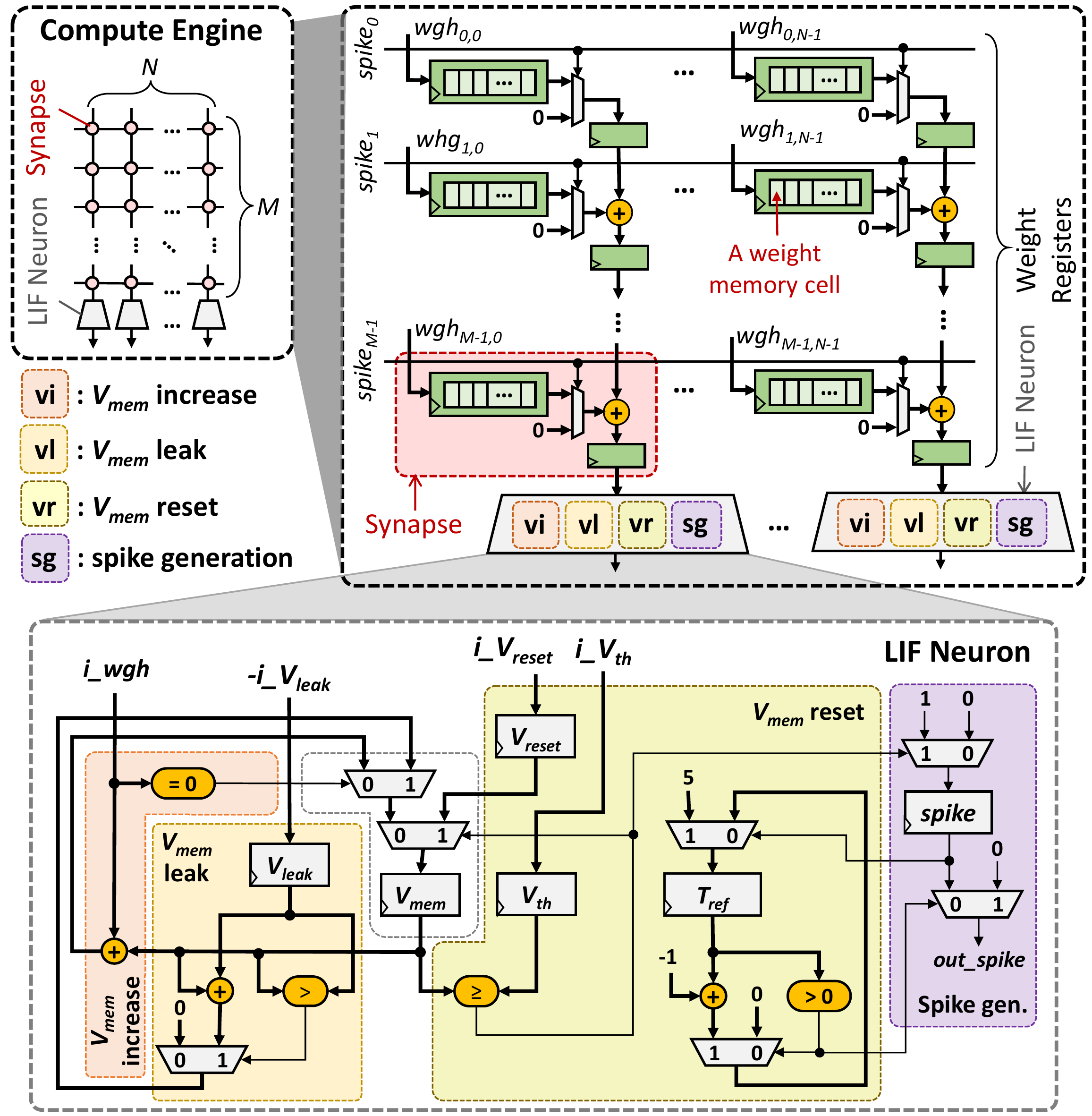}
\vspace{-0.7cm}
\caption{The detailed architecture of SNN compute engine, showing the weight registers and neuron operations.}
\label{Fig_SNNacc_Engine}
\vspace{-0.4cm}
\end{figure}

\textbf{SNN Accelerator Architecture:}
We consider the typical architectures of SNN accelerators and compute engine as shown in Fig.~\ref{Fig_SNNacc_Overview} and Fig.~\ref{Fig_SNNacc_Engine}, respectively, based on the design of~\cite{Ref_Frenkel_ODIN_TBCAS19}.
Here, we focus on the compute engine, as it is responsible for generating output spikes and determining accuracy. 
The compute engine has a synapse crossbar, and each synapse hardware employs a register to store a weight value. 
We consider 8-bit precision for each weight as it has a good accuracy-memory trade-off~\cite{Ref_Putra_FSpiNN_TCAD20}.
To optimize the routing (area), each synapse employs an adder to sum its weight value with an accumulated value from the previous synapses in the same column, hence a neuron only needs a single input from the connected synapses.
In each neuron hardware, the membrane potential ($V_{mem}$) is increased by the weight ($wgh$) each time a spike arrives, otherwise, the $V_{mem}$ is decreased. 
If the $V_{mem}$ reaches the threshold potential ($V_{th}$), a spike is produced. 
Then, the $V_{mem}$ resets to $V_{reset}$, and the neuron does not generate spikes for some time, known as refractory period ($T_{ref}$).
Hence, a LIF neuron model has four main operations: (1) $V_{mem}$ increase, (2) $V_{mem}$ leak, (3) $V_{mem}$ reset, and (4) spike generation, as shown in Fig.~\ref{Fig_SNNacc_Overview}. 
The membrane potential dynamics are defined by operations (1)-(3).

%%%%%%%%%%%%%%%%%%%%%%
\vspace{-0.1cm}
\subsection{Transient Fault Modeling}
\label{Sec_Prelim_FaultModel}

The SNN compute engine consists of synapse and neuron parts, each having specialized hardware circuitry. 
Therefore, we need to define the transient fault modeling for each part.

\begin{figure}[t]
\centering
\includegraphics[width=\linewidth]{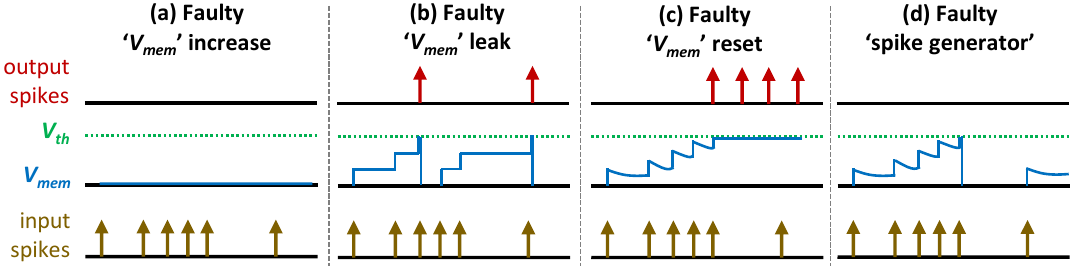}
\vspace{-0.7cm}
\caption{Overview of different faulty neuron operations.}
\label{Fig_FaultyNeuronOps}
\vspace{-0.5cm}
\end{figure}

\begin{itemize}[leftmargin=*]
    \item \textbf{Synapse Part:} 
    A fault in a synapse hardware only affects a single weight bit in form of a bit flip. 
    This faulty bit persists until it is overwritten with a new bit value.
    \item \textbf{Neuron Part:} 
    Soft errors in the neuron hardware can manifest in different forms depending upon the type of operation being executed on the neuron hardware, as discussed in the following (see overview in Fig.~\ref{Fig_FaultyNeuronOps}).
    \begin{itemize}
        \item \textit{Soft errors in the `$V_{mem}$ increase' operation} make the neuron unable to increase $V_{mem}$, hence this neuron is unable to reach $V_{th}$ and does not produce any spikes. 
        \item \textit{Soft errors in the `$V_{mem}$ leak' operation} make the neuron unable to decrease $V_{mem}$. 
        \item \textit{Soft errors in the `$V_{mem}$ reset' operation} make the neuron unable to reset $V_{mem}$, hence this neuron continuously produces spikes.
        \item \textit{Soft errors in the `spike generation'} make the neuron unable to produce any spikes.
        \item These faulty operations persist until the neuron parameters are replaced with a new set of parameters.
    \end{itemize}
\end{itemize}
 
\begin{figure}[hbtp]
\vspace{-0.4cm}
\centering
\includegraphics[width=\linewidth]{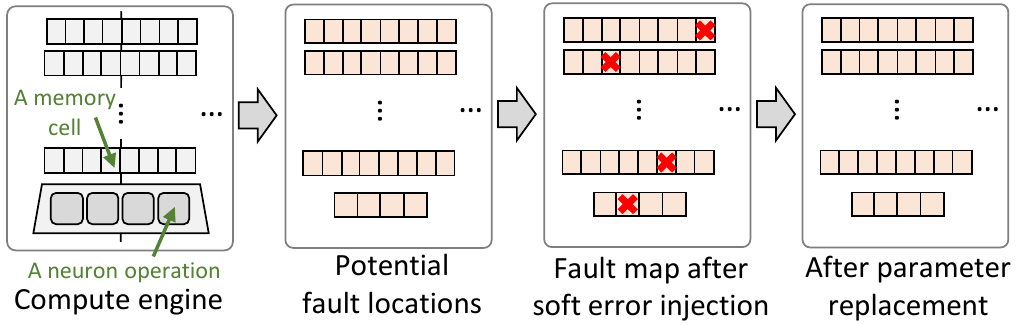}
\vspace{-0.7cm}
\caption{Our steps for conducting soft error generation and distribution on the SNN compute engine.}
\label{Fig_FaultMap}
\vspace{-0.5cm}
\end{figure}

\textbf{Soft Error Generation and Distribution:} 
Soft errors typically occur in random locations of a chip~\cite{Ref_Baumann_SoftErrors_DnT05}, leading to a certain fault map. 
Following are the steps for generating and distributing soft errors (see overview in Fig.~\ref{Fig_FaultMap}). 
\begin{itemize}[leftmargin=*]
    \item We consider each weight memory cell and neuron operation as the potential fault locations. 
    \item We generate soft errors considering the given fault rate, and distribute them randomly across the potential fault locations. 
    \item If a fault occurs in a memory cell, we flip the stored bit, which persists until it is overwritten by a new value.
    If an error occurs in a neuron operation, we randomly select the type of faulty operation, which persists until new parameters are set for the respective neuron.
\end{itemize}

%%%%%%%%%%%%%%%%%%%%%%%%%%%%%%%%%%%%%%%%%%%%%%%%%%
%%%%%%%%%%%%%%%%%%%%%%%%%%%%%%%%%%%%%%%%%%%%%%%%%%
\section{SoftSNN Methodology}
\label{Sec_SoftSNN}
\vspace{-0.1cm}

\begin{figure*}[t]
\centering
\includegraphics[width=\linewidth]{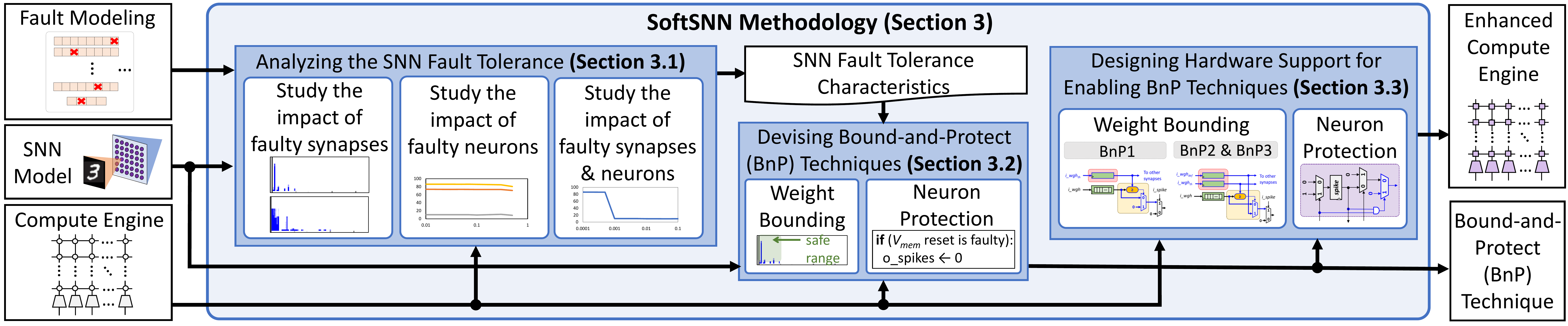}
\vspace{-0.7cm}
\caption{Overview of our SoftSNN methodology, and its key steps for mitigating soft errors in the SNN compute engine.}
\label{Fig_SoftSNN_Method}
\vspace{-0.5cm}
\end{figure*}

The overview of SoftSNN methodology is shown in Fig.~\ref{Fig_SoftSNN_Method}, and the detailed discussion of its key steps is provided in Section~\ref{Sec_SoftSNN_FTAnalysis}-\ref{Sec_SoftSNN_HWBnP}. 

%%%%%%%%%%%%%%%%%%%%%%
\vspace{-0.2cm}
\subsection{SNN Fault Tolerance Analysis}
\label{Sec_SoftSNN_FTAnalysis}

\begin{figure}[hbtp]
\vspace{-0.4cm}
\centering
\includegraphics[width=\linewidth]{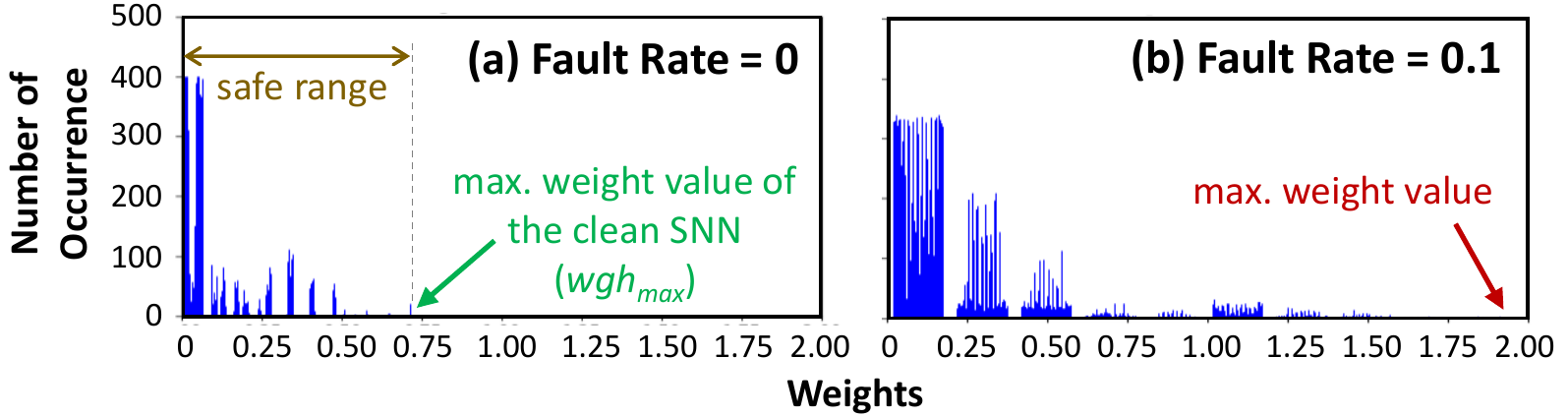}
\vspace{-0.7cm}
\caption{Soft errors can increase weight values, which may surpass the maximum weight value (i.e., $wgh_{max}$) from pre-trained SNN without soft errors (clean SNN).}
\label{Fig_Observe_FaultyWreg_Wdist}
\vspace{-0.4cm}
\end{figure}

\begin{figure}[hbtp]
\vspace{-0.3cm}
\centering
\includegraphics[width=\linewidth]{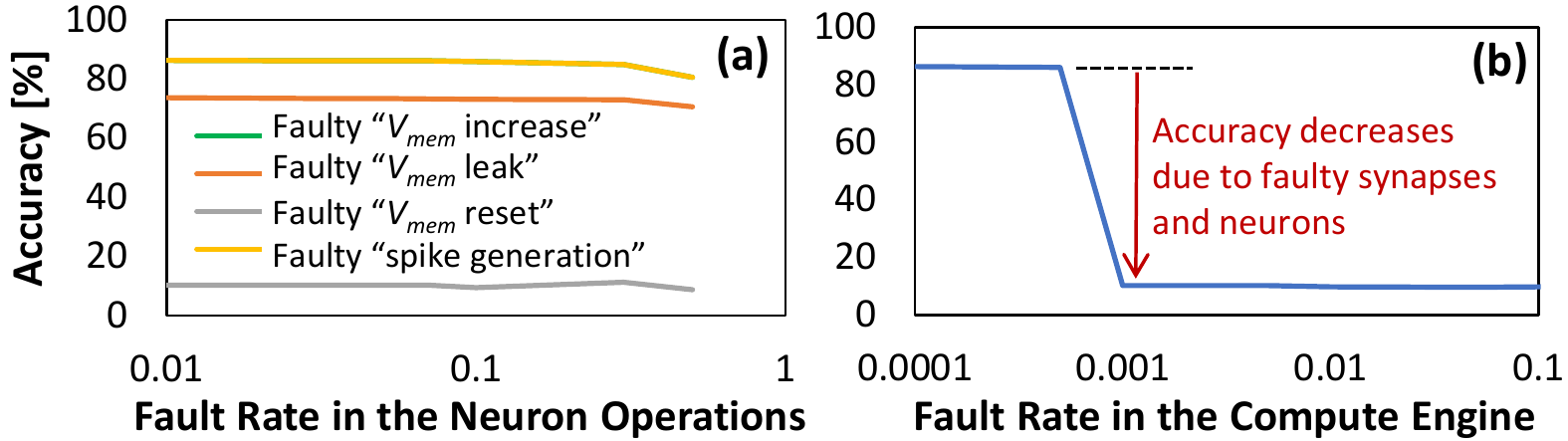}
\vspace{-0.7cm}
\caption{The impact of (a) faulty neuron operations, and (b) faulty weight registers and neuron operations, on accuracy.}
\label{Fig_Observe_FaultyNeuronsCE}
\vspace{-0.5cm}
\end{figure}

\textit{Our SoftSNN first performs SNN fault tolerance analysis by characterizing the behavior of a given SNN model under different soft error profiles for the underlying hardware}, which provides beneficial information for devising lightweight soft error mitigation techniques. 
To do this, we perform experimental case studies for a 400-neuron network with MNIST~\footnote{We use the MNIST dataset as its size enables fast exploration for SNN fault tolerance analysis. Moreover, each SNN input has the same time range and coding for its spike train representation, and the employed STDP learning limits the weights in a certain range of positive values, thereby making any workloads representative for the analysis.} as described in the following.
\begin{itemize}[leftmargin=*]
    \item \textbf{Impact of faulty synapses:}
    We inject soft errors on the weight registers by randomly flipping the stored weight bits. 
    The experimental results are shown in Figs.~\ref{Fig_Observe_FaultyWreg} and ~\ref{Fig_Observe_FaultyWreg_Wdist}, from which we derive the following observations. 
    \begin{itemize}
        \item Soft errors may increase or decrease weight values, and the increased ones have a more severe impact on accuracy since they trigger the neurons to generate spikes more frequently, thereby dominating classification.
        \item Increased weights may be recognized by employing the maximum weight value ($wgh_{max}$) of the pre-trained SNN without soft errors (clean SNN) as a threshold.   
    \end{itemize}
\end{itemize}
\begin{itemize}[leftmargin=*]
    \item \textbf{Impact of faulty neurons:}
    We inject soft errors on the neurons by randomly generating faulty neuron operations. 
    The experimental results are shown in Fig.~\ref{Fig_Observe_FaultyNeuronsCE}(a), from which we obtain the following observations.
    \begin{itemize}
        \item Inference with faulty ‘$V_{mem}$ increase’, ‘$V_{mem}$ leak’, and ‘spike generation’ can achieve tolerable accuracy, as their faulty behavior does not make the neurons dominate classification, and the function for classifying the same input class may be substituted by other (non-faulty) neurons. 
        Therefore, \textit{these faulty neurons can still be employed for SNN processing.}
        \item Inference with faulty ‘$V_{mem}$ reset’ can decrease the accuracy significantly, as this faulty behavior makes the neurons' membrane potential stays greater or equal to the threshold potential ($V_{mem} \geq V_{th}$), thereby generating (faulty) burst spikes and dominating classification. 
        Therefore, \textit{these faulty neurons should not be employed for SNN processing.}
    \end{itemize}
    \item \textbf{Impact of faulty synapses and neurons:}
    We inject soft errors by randomly flipping bits on the weight registers and generate faulty neuron operations. 
    The experimental results in Fig.~\ref{Fig_Observe_FaultyNeuronsCE}(b) show that the faulty compute engine can severely decrease accuracy, thereby emphasizing the importance of soft error mitigation. 
\end{itemize}

%%%%%%%%%%%%%%%%%%%%%%
\subsection{Our Bound-and-Protect (BnP) Techniques}
\label{Sec_SoftSNN_BnPTech}

To detect and mitigate soft errors at run time, \textit{our SoftSNN methodology employs the Bound-and-Protect (BnP) technique, which bounds the weight values within a safe range that does not make neurons hyper-active (i.e., \textbf{weight bounding}), and protects the neurons from performing faulty operations that can significantly decrease accuracy (i.e., \textbf{neuron protection})}.    

\textbf{Weight Bounding:}
This mechanism clips the weight values that are greater or equal to the weight threshold ($wgh \geq wgh_{th}$), and replaces them with a pre-defined value ($wgh_{def}$), as stated in Eq.~\ref{Eq_WeightBounding}. 
Hence, each weight has the bounding value ($wgh_{b}$) that does not trigger neurons' hyper-activity.  
To define $wgh_{th}$, we leverage the SNN fault tolerance characteristics from Section~\ref{Sec_SoftSNN_FTAnalysis}. 
We consider the range of weight values from the pre-trained SNN without soft errors (clean SNN) as the \textit{safe range}, and employ its maximum value as the weight threshold ($wgh_{th} = wgh_{max}$), as shown in Fig.~\ref{Fig_Observe_FaultyWreg_Wdist}(a). 

\vspace{-0.3cm}
\begin{equation}
\small
 wgh_{b} = 
\begin{cases}
wgh_{def} & \text{if} \; wgh \geq wgh_{th} \\ 
wgh & \text{if} \; \text{otherwise}
\end{cases}
\label{Eq_WeightBounding}
\end{equation}

\textbf{Neuron Protection:}
This mechanism focuses on mitigating faulty `$V_{mem}$ reset' operations, as suggested by analysis in Section~\ref{Sec_SoftSNN_FTAnalysis}.
We detect the faulty `$V_{mem}$ reset' operation in each neuron by monitoring the comparison output of $V_{mem}$ $\geq$ $V_{th}$. 
If the output is `true' for multiple clock cycles (e.g., $\geq$ 2 clock cycles in this work), then it indicates that the `$V_{mem}$ reset' operation does not work properly.
To efficiently address this, we disable the spike generation to prevent the corresponding neuron from generating burst spikes.

\textit{We leverage these mechanisms for devising three variants of BnP techniques (i.e., BnP1, BnP2, and BnP3)}, which provide trade-offs in terms of accuracy, latency, and energy for soft error mitigation. 
\begin{itemize}[leftmargin=*]
    \item \textbf{BnP1 Technique:}
    It replaces the weights that are greater or equal to the $wgh_{th}$ with zero. 
    Therefore, the BnP1 can be stated as Eq.~\ref{Eq_WeightBounding} with $wgh_{def} = 0$. 
    \item \textbf{BnP2 Technique:}
    It replaces the weights that are greater or equal to the $wgh_{th}$ with the maximum weight value from clean SNN ($wgh_{max}$). 
    Therefore, the BnP2 can be stated as Eq.~\ref{Eq_WeightBounding} with $wgh_{def} = wgh_{max}$.
    \item \textbf{BnP3 Technique:}
    It replaces the weights that are greater or equal to the $wgh_{th}$ with a highly probable value from the weight distribution of clean SNN ($wgh_{hp}$). 
    Therefore, the BnP3 can be stated as Eq.~\ref{Eq_WeightBounding} with $wgh_{def} = wgh_{hp}$.
    \item \textbf{For All BnP Techniques:} 
    We continuously monitor the neuron dynamics, and if the faulty `$V_{mem}$ reset' operation occurs, we disable the respective spike generation. 
\end{itemize}

%%%%%%%%%%%%%%%%%%%%%%
\vspace{-0.2cm}
\subsection{Hardware Support for BnP Techniques}
\label{Sec_SoftSNN_HWBnP}

Performing the BnP techniques on the SNN accelerators at run time is challenging, as these accelerators typically have fixed dataflows.
Hence, \textit{we propose lightweight self-healing hardware enhancements to support the deployment of our BnP techniques on the SNN accelerators without changing the dataflows}, as described in the following.

\begin{figure}[hbtp]
\vspace{-0.3cm}
\centering
\includegraphics[width=\linewidth]{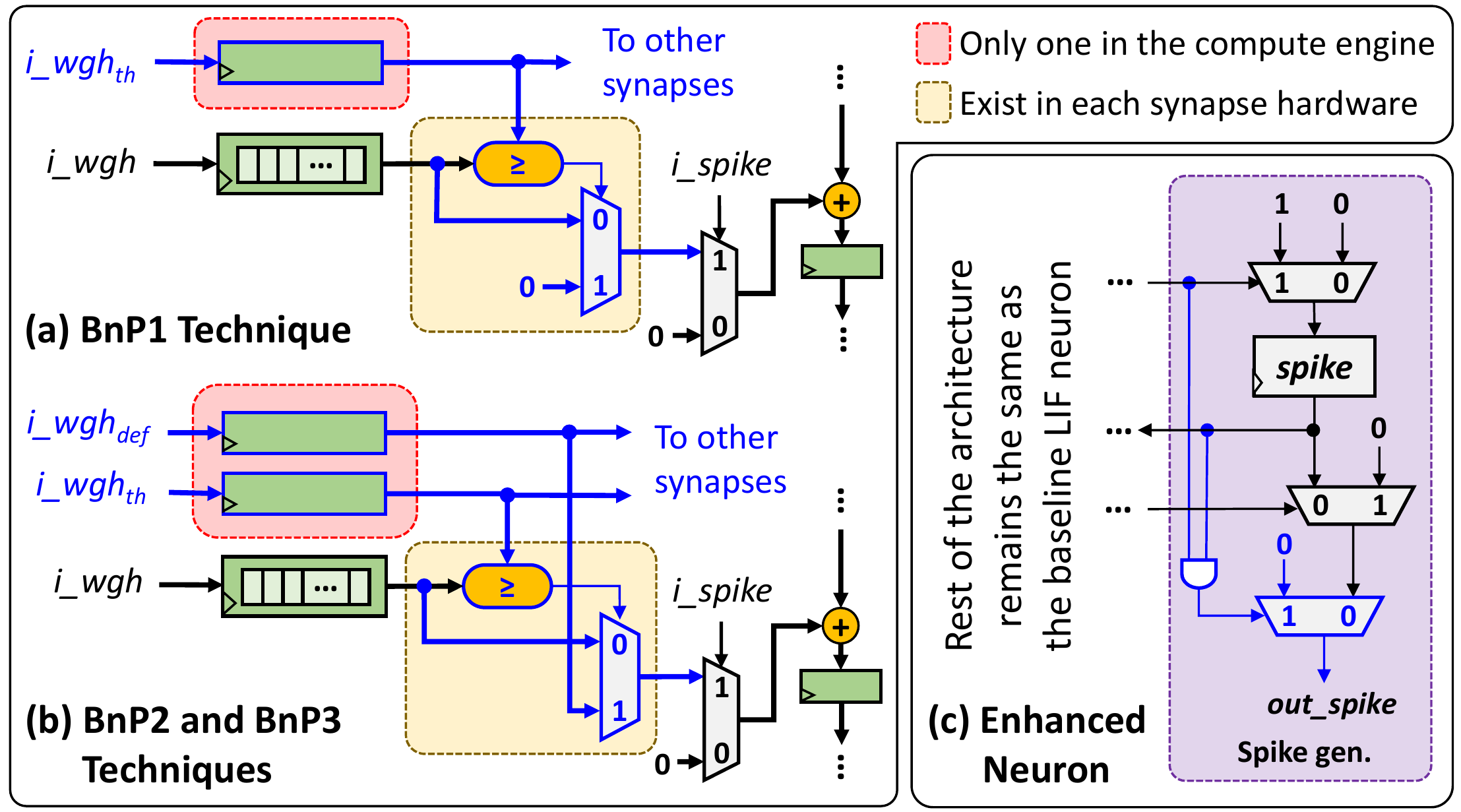}
\vspace{-0.7cm}
\caption{The proposed synapse architectures for (a) BnP1, and (b) BnP2 and BnP3 techniques. (c) The proposed neuron architecture for mitigating faulty `$V_{mem}$ reset' operation. All circuit enhancements are highlighted in blue.}
\label{Fig_SoftSNN_HWenhance}
\vspace{-0.4cm}
\end{figure}

\textbf{Synapse Part:}
The synapse enhancements aim at enabling the weight bounding, and they depend on the type of BnP technique.
\begin{itemize}[leftmargin=*]
    \item In case of the \textbf{BnP1 Technique}, we add (1) a radiation-hardened register for storing the weight threshold $wgh_{th}$, which is used for all synapses in the compute engine, and (2) the hardened combinational logic units for performing a comparison and multiplexing in each synapse; see Fig.~\ref{Fig_SoftSNN_HWenhance}(a). 
    \item In case of the \textbf{BnP2 and BnP3 Techniques}, we add (1) two radiation-hardened registers for storing the weight threshold $wgh_{th}$ and the pre-defined weight value $wgh_{def}$ respectively, which are used for all synapses in the compute engine, and (2) the hardened combinational logic units for performing a comparison and multiplexing in each synapse; see Fig.~\ref{Fig_SoftSNN_HWenhance}(b). 
\end{itemize}

\textbf{Neuron Part:}
To recognize faulty `$V_{mem}$ reset' operation, we monitor the comparison output of $V_{mem} \geq V_{th}$.  
If the output is `true' for $\geq$ 2 clock cycles, then the `$V_{mem}$ reset' operation is faulty. 
To ensure that such faulty operations do not result in burst spikes, we add an AND logic and a multiplexer to leverage the current and upcoming outputs for determining if the neuron should generate a spike in the next cycle, as shown in Fig.~\ref{Fig_SoftSNN_HWenhance}(c).  

\textbf{Radiation Hardening:}
Since our hardware enhancements can also be affected by soft errors, we consider radiation hardened components for all the new hardware extensions to make them resistant to high-energy particle strikes. 
To do this, the hardening techniques that improve the fabrication process (e.g., re-sizing transistor and insulating substrates~\cite{Ref_Garg_RadHard_TVLSI09, Ref_Huang_RadHard_Nuclear19}) are employed.
We only need to harden the additional components, since they will provide correct values which can replace the corrupted bits in the subsequent circuits. 
Hence, the overhead of the hardening process is relatively low as compared to the full architecture of the SNN hardware, and will be discussed further in Section~\ref{Sec_Results_Overheads} (i.e., area overhead).

%%%%%%%%%%%%%%%%%%%%%%%%%%%%%%%%%%%%%%%%%%%%%%%%%%
%%%%%%%%%%%%%%%%%%%%%%%%%%%%%%%%%%%%%%%%%%%%%%%%%%
\vspace{-0.2cm}
\section{Evaluation Methodology}
\label{Sec_Evaluation}

\begin{figure}[hbtp]
\centering
\includegraphics[width=0.9\linewidth]{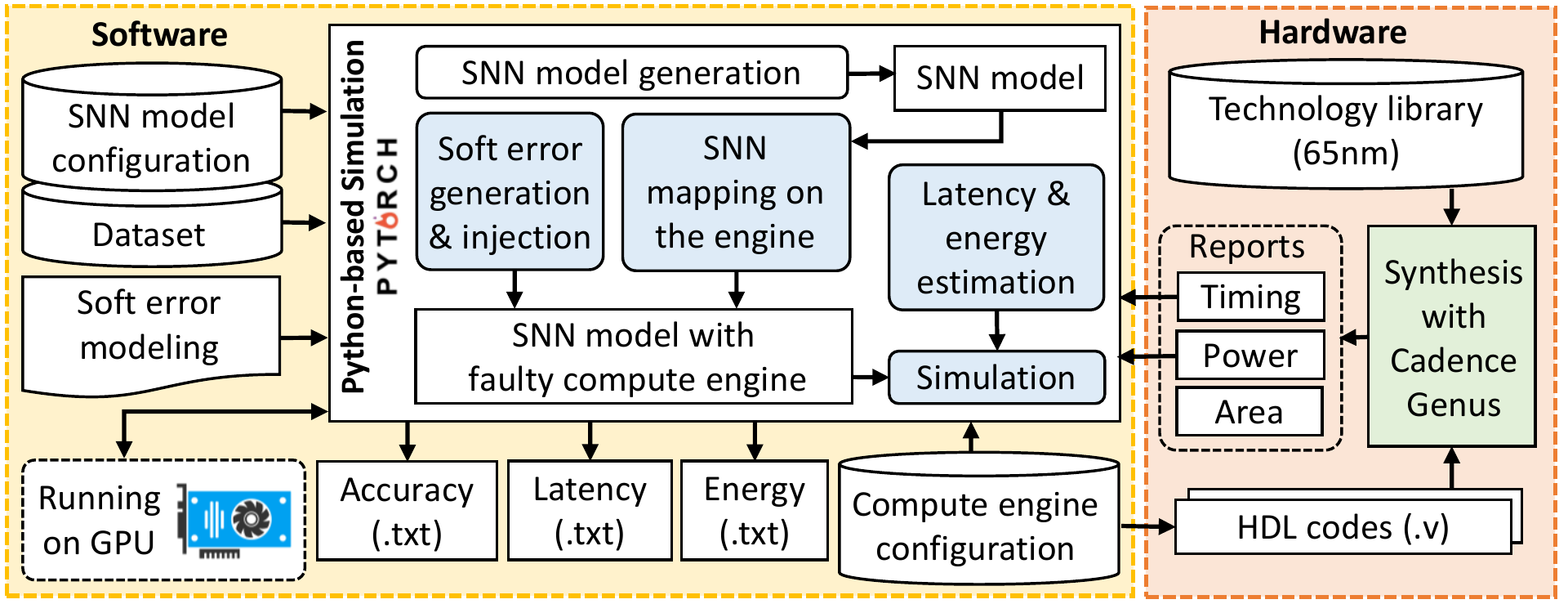}
\vspace{-0.4cm}
\caption{The experimental setup and tools flow.}
\label{Fig_Eval_Method}
\vspace{-0.5cm}
\end{figure}

We deploy the experimental setup presented in Fig.~\ref{Fig_Eval_Method} for evaluating our SoftSNN methodology, and \textit{adopt the same evaluation conditions as employed widely by the SNN community}.
We use the fully-connected network as shown in Fig.~\ref{Fig_SNN_SoftErrors}(a) with a different number of neurons for evaluating the generality of our SoftSNN. 
For simplicity, we refer a network with $i$-number of neurons to as N$i$. 
We employ the MNIST and Fashion MNIST datasets as workloads. 
For comparison partners, we consider (1) the SNN without mitigation (i.e., No Mitigation), and (2) the SNN with 3x redundant executions and majority voting (i.e., Re-execution in TMR mode).

\textbf{Accuracy Evaluation:}
We use a Python-based framework~\cite{Ref_Hazan_BindsNET_FNINF18}, which run on multi-GPU machines with Nvidia RTX 2080 Ti, while simulating the SNN accelerator architecture of Fig.~\ref{Fig_SNNacc_Overview}.

\textbf{Hardware Evaluations:}
We implement the SNN compute engine shown in Fig.~\ref{Fig_SNNacc_Engine} with a 256x256 synapse crossbar.
Its timing, power, and area are obtained through hardware synthesis using the Cadence Genus with a 65nm CMOS technology library.
We estimate the latency of compute engine \textit{for both with and without hardware enhancements}, by leveraging the computation time for an inference of a single input.
Afterward, we leverage the obtained latency and power to estimate the energy consumption of the compute engine.

%%%%%%%%%%%%%%%%%%%%%%%%%%%%%%%%%%%%%%%%%%%%%%%%%%
%%%%%%%%%%%%%%%%%%%%%%%%%%%%%%%%%%%%%%%%%%%%%%%%%%
\vspace{-0.1cm}
\section{Results and Discussion}
\label{Sec_Results}
\vspace{-0.1cm}

\begin{figure*}[t]
\centering
\includegraphics[width=0.92\linewidth]{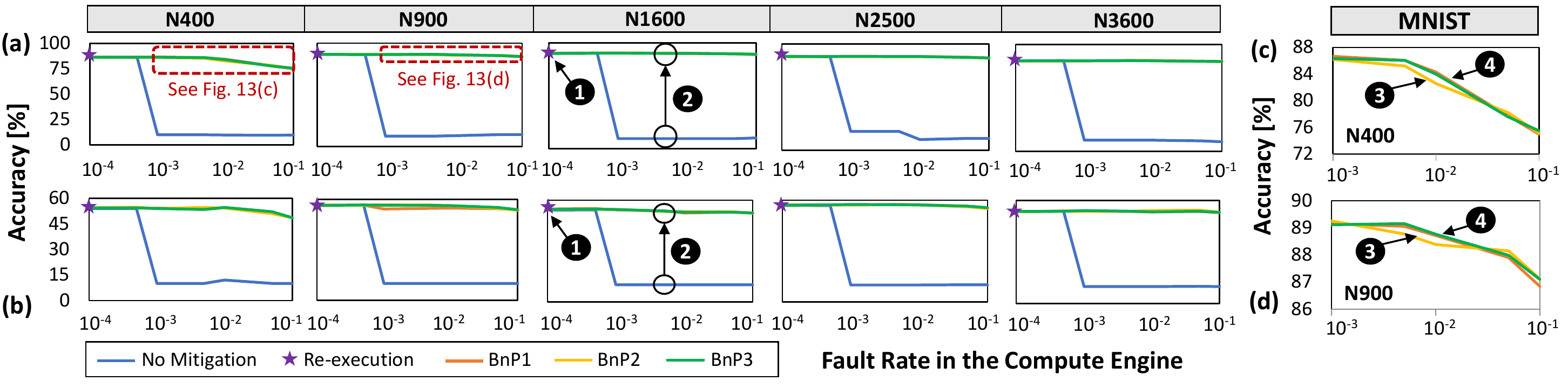}
\vspace{-0.4cm}
\caption{Results on accuracy across different mitigation techniques, network sizes, fault rates, and workloads: (a) MNIST and (b) Fashion MNIST. Detailed accuracy profiles on MNIST are shown for (c) N400 and (d) N900.}
\label{Fig_Results_Accuracy}
\vspace{-0.2cm}
\end{figure*}

\begin{figure*}[t]
\vspace{-0.2cm}
\centering
\includegraphics[width=0.92\linewidth]{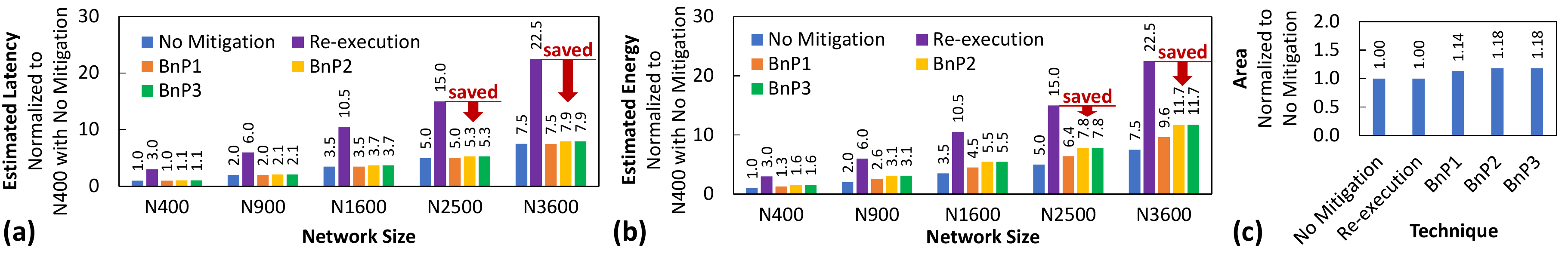}
\vspace{-0.4cm}
\caption{Comparisons across different techniques and network sizes on (a) latency and (b) energy for an inference of a single input, and (c) area. Results for MNIST and Fashion MNIST are similar, as these workloads have the same input dimension.}
\label{Fig_Results_Overheads}
\vspace{-0.5cm}
\end{figure*}

%%%%%%%%%%%%%%%%%%%%%%
\subsection{Accuracy Comparisons}
\label{Sec_Results_Accuracy}
\vspace{-0.1cm}

Fig.~\ref{Fig_Results_Accuracy} presents the experimental results for the accuracy of different mitigation techniques across various test scenarios. 
The re-execution technique can achieve high accuracy as shown by \circledB{1}, since it employs redundant executions to ensure consistent outputs, which indicates that the executions are minimally affected by soft errors.
Meanwhile, our BnP techniques (BnP1, BnP2, and BnP3) achieve comparable accuracy to the re-execution, and significantly improve accuracy compared to the SNN without mitigation, i.e., by up to 80\% and 47\% for MNIST and Fashion MNIST, respectively; see \circledB{2}.
The reason is that, our techniques employ safe weight values and neuron operations to avoid faulty neural dynamics that significantly decrease accuracy.    
We observe that the BnP2 has slightly lower accuracy compared to the BnP1 and the BnP3, as it employs $wgh_{max}$ as $wgh_{def}$, whose values have low probability in the weight distribution of clean SNN; see \circledB{3}.
Hence, the generated neural dynamics do not closely match the neural dynamics of clean SNN. 
We also observe that the BnP1 and the BnP3 have comparable accuracy, as their $wgh_{def}$ are relatively close to each other in the weight distribution of clean SNN; see \circledB{4}. 
Hence, the neural dynamics of the BnP1 and the BnP3 are similar.
However, the BnP3 has better applicability for diverse applications than the BnP1, since the $wgh_{def}$ in the BnP3 can be updated for different weight distributions.
These results show that \textit{our BnP techniques are effective for mitigating soft errors in the SNN compute engine at run time without re-execution.} 

%%%%%%%%%%%%%%%%%%%%%%
\vspace{-0.2cm}
\subsection{Latency, Energy, and Area Overheads}
\label{Sec_Results_Overheads}
\vspace{-0.1cm}

Besides accuracy, we also evaluate the design overheads (i.e., latency, energy, and area) incurred by different mitigation techniques.  

\textbf{Latency:}
Experimental results for latency are shown in Fig.~\ref{Fig_Results_Overheads}(a).
We observe that the re-execution technique incurs $\sim$3x latency compared to the SNN without mitigation, as it employs redundant executions for loading parameters on the compute engine and performing neural operations.
Meanwhile, our BnP techniques only incur less than 1.06x latency compared to the SNN without mitigation, and reduce latency by up to 3x compared to the re-execution, due to our efficient hardware modifications that minimally affect the latency (i.e., a small number of registers and/or combinational logic units without noticeably affecting the critical path).
In this manner, our BnP-enhanced compute engine preserves the existing processing dataflow, and enables reliable SNN executions for latency-constrained (real-time) applications.

\textbf{Energy Consumption:}
Experimental results for energy consumption are shown in Fig.~\ref{Fig_Results_Overheads}(b).
We observe that, the re-execution technique incurs 3x energy overhead compared to the SNN without mitigation, due to its redundant executions.
Meanwhile, our BnP techniques incur less than 1.6x energy consumption when compared to the SNN without mitigation, and reduce the energy consumption by up to 2.3x as compared to the re-execution. 
Note, compared to the original hardware executing SNN without mitigation, the slight increase in the energy consumption of our BnP-enhanced hardware is due to the additional hardware components to enable reliable SNN execution without incurring noticeable latency/performance overheads.
Moreover, since the redundant executions are completely avoided, our techniques substantially optimize the energy consumption as compared to the re-execution based mitigation technique.

\textbf{Area:}
Experimental results for area are shown in Fig.~\ref{Fig_Results_Overheads}(c).
We observe that our BnP-enhanced compute engine incurs tolerable area overhead (i.e., 14\% for the BnP1, and 18\% for the BnP2 and the BnP3) as compared to the compute engine without enhancements. 
These area overheads mainly come from additional components in synapses, as the synapse crossbar dominates the area of compute engine.  
Furthermore, the area overhead also represents the cost of the new radiation-hardened components to ensure reliable SNN execution. 
Note, we only need to harden the additional components for providing correct bits to the subsequent circuits, thereby correcting the corrupted bits and ensuring reliable executions in the respective circuits with low overhead.

\vspace{0.1cm}
In summary, all these results show that \textit{our BnP techniques effectively mitigate soft errors, while significantly reducing the latency and the energy of SNN executions as compared to the re-execution based mitigation technique.} 

%%%%%%%%%%%%%%%%%%%%%%%%%%%%%%%%%%%%%%%%%%%%%%%%%%
%%%%%%%%%%%%%%%%%%%%%%%%%%%%%%%%%%%%%%%%%%%%%%%%%%
\vspace{-0.1cm}
\section{Conclusion}
\label{Sec_Conclusion}

We propose the SoftSNN methodology for mitigating soft errors in SNN accelerators without re-execution. 
Our SoftSNN analyzes the SNN characteristics under soft errors, performs weight bounding and neuron protection, and devises efficient hardware enhancements to enable the proposed technique. 
The results show that, our SoftSNN maintains high accuracy while reducing latency and energy, compared to the re-execution based mitigation technique, thereby enabling reliable SNN executions for real-time and energy-efficient applications. 

%%%%%%%%%%%%%%%%%%%%%%%%%%%%%%%%%%%%%%%%%%%%%%%%%%
%%%%%%%%%%%%%%%%%%%%%%%%%%%%%%%%%%%%%%%%%%%%%%%%%%
\vspace{-0.1cm} 
\begin{acks}
This work was partly supported by Intel Corporation through Gift funding for the project ``Cost-Effective Dependability for Deep Neural Networks and Spiking Neural Networks”, and by Indonesia Endowment Fund for Education (LPDP). This work was also jointly supported by the NYUAD Center for Interacting Urban Networks (CITIES), funded by Tamkeen under the NYUAD Research Institute Award CG001 and Center for CyberSecurity (CCS), funded by Tamkeen under the NYUAD Research Institute Award G1104.
\end{acks}

%%%%%%%%%%%%%%%%%%%%%%%%%%%%%%%%%%%%%%%%%%%%%%%%%%
%%%%%%%%%%%%%%%%%%%%%%%%%%%%%%%%%%%%%%%%%%%%%%%%%%
\vspace{-0.2cm}
\bibliographystyle{ACM-Reference-Format}
\bibliography{bibliography}

\end{spacing}
\end{document}